\documentclass[preprint2,twoside]{hwo}

\usepackage{graphicx}
\usepackage{tabularx,ragged2e,booktabs,babel}
\newcolumntype{L}{>{\RaggedRight\arraybackslash}X}
\newcolumntype{C}[1]{>{\Centering\arraybackslash%
        \hsize=#1\hsize\linewidth=\hsize}X}

\input{hwo.h}

\begin{document}

\title{\textbf{\LARGE Identifying rocky planets and water worlds among sub-Neptune-sized exoplanets with the Habitable Worlds Observatory}}

\author {\textbf{\large 
Renyu Hu,$^1$ Michiel Min,$^2$ Max Millar-Blanchaer,$^3$ Jacob Lustig-Yaeger,$^4$ Tyler Robinson$^5$}}
\affil{$^1$\small\it Jet Propulsion Laboratory, California Institute of Technology, Pasadena, California, USA \email{renyu.hu@jpl.nasa.gov}}
\affil{$^2$\small\it SRON Netherlands Institute for Space Research, Leiden, The Netherlands}
\affil{$^3$\small\it Department of Physics, University of California, Santa Barbara, Santa Barbara, California, USA}
\affil{$^4$\small\it John Hopkins Applied Physics Laboratory, Laurel, Maryland, USA}
\affil{$^5$\small\it Lunar and Planetary Laboratory, University of Arizona, Tucson, Arizona, USA}

\author{\small{\bf Contributing Authors:} Jennifer Burt (JPL), Athena Coustenis (Paris Observatory), Mario Damiano (JPL), Chuanfei Dong (Boston University), Courtney Dressing (UC Berkeley), Luca Fossati (Austrian Academy of Sciences), Stephen Kane (UC Riverside), Soumil Kelkar (University of Groningen), Tim Lichtenberg (University of Groningen), Jean-Baptiste Ruffio (UC San Diego), Dibyendu Sur (Catholic Univ. of America and NASA GSFC), Armen Tokadjian (JPL), Martin Turbet (CNRS).
}

\author{\footnotesize{\bf Endorsed by:}
Munazza Alam (STScI), Eleonora Alei (NASA GSFC), Natalie Allen (Johns Hopkins University), Narsireddy Anugu (Georgia State University), Komal Bali (ETH Zurich), Katherine Bennett (Johns Hopkins University), Alan Boss (Carnegie Earth and Planets Lab), Kara Brugman (Arizona State University), Douglas Caldwell (SETI Institute), Aarynn Carter (STScI), Arnot David (The Open University), Jessie Christiansen (Caltech/IPAC), Nicolas Crouzet (Kapteyn Astronomical Institute), Diana Dragomir (University of New Mexico), Megan Gialluca (University of Washington), Christopher Glein (Southwest Research Institute), Kenneth Goodis Gordon (University of Central Florida), Caleb Harada (UC Berkeley), Natalie Hinkel (Louisiana State University), Theodora Karalidi (University of Central Florida), Preethi Karpoor (Indian Institute of Astrophysics), Finnegan Keller (Arizona State University), Eliza Kempton (University of Chicago), Joshua Krissansen-Totton (University of Washington), Alen Kuriakose (KU Leuven), Adam Langeveld (Johns Hopkins University), Eunjeong Lee (EisKosmos), Briley Lewis (UC Santa Barbara), Mercedes López-Morales (STScI), David Montes (Universidad Complutense de Madrid), Arnaud Salvador (German Aerospace Center DLR), Gaetano Scandariato (INAF), Edward Schwieterman (UC Riverside), Melinda Soares-Furtado (UW Madison), Johanna Teske (Carnegie Earth and Planets Lab), Thaddeus Komacek (University of Oxford), Margaret Turcotte Seavey (University of Maryland), Vincent Van Eylen (UCL), Hannah Wakeford (University of Bristol), Lauren Weiss (University of Notre Dame), Thomas Wilson (University of Warwick), Nicholas Wogan (NASA Ames).
}



\begin{abstract}
Astronomers are debating whether the plentiful ``sub-Neptune'' exoplanets -- worlds a bit larger than Earth but smaller than Neptune -- are predominantly rocky planets, water-rich ``ocean worlds,'' or gas-enshrouded mini-Neptunes. This question is crucial because such sub-Neptune-sized planets are among the most common in our galaxy, yet we have no analog in our own solar system, making them a key to understanding planet formation and diversity. It also directly impacts the search for habitable worlds: larger-than-Earth planets with solid surfaces or oceans could support life, whereas gas-rich mini-Neptunes likely cannot. However, distinguishing these types using only a planet's mass and radius is very challenging, because different compositions can produce similar densities, leaving a world's nature ambiguous with current data. The proposed Habitable Worlds Observatory (HWO), a future NASA flagship telescope, offers a solution. HWO could directly image and spectroscopically analyze starlight reflected from $50\sim100$ sub-Neptunes around nearby stars, aiming to reveal their atmospheric compositions and potential surfaces. Using visible and near-infrared spectroscopy along with sensitive polarimetry, HWO would detect atmospheric gases (such as water vapor, methane, and carbon dioxide) and search for telltale surface signatures, including rock absorption features and the characteristic reflectivity patterns of oceans. By analyzing these signals, we could determine whether sub-Neptunes are large rocky planets or water worlds rather than gas-dominated mini-Neptunes. Crucially, expanding the search beyond Earth-sized planets to include these abundant sub-Neptunes may uncover entirely new classes of potentially habitable worlds, directly advancing HWO's mission to identify and characterize planets that could support life.
\\
\end{abstract}

\vspace{2cm}

\section{Science Goal}

The fundamental question we aim to address in this HWO science case is: \textbf{At what frequency do nearby stars host rocky planets or water worlds versus mini Neptunes with gaseous envelopes?}

\begin{figure}[ht!]
    \centering
    \includegraphics[width=0.5\textwidth]{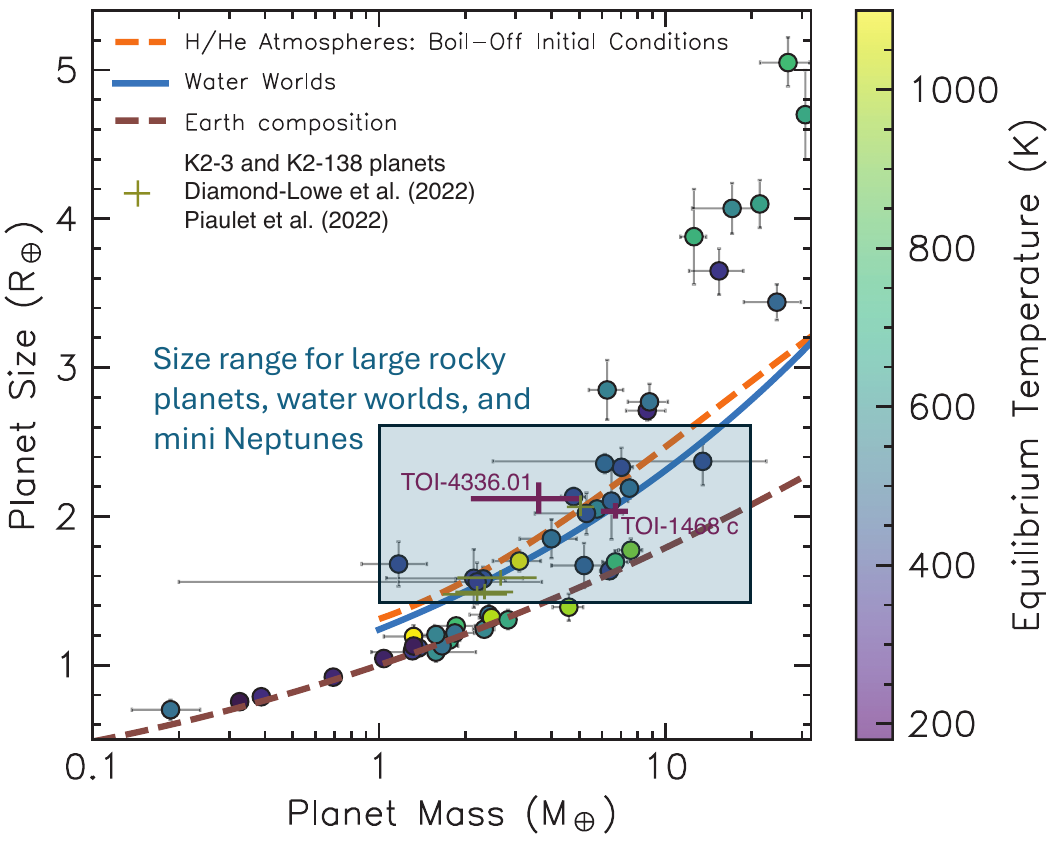}
    \caption{Diverse density and composition of small exoplanets. The plot shows M-dwarf planets that have precision mass and radius measurements \citep{luque2022density}. Small planets of FGK stars also occupy a broad parameter space that spans from Earth composition to water worlds or planets with massive H/He envelopes \citep{parc2024super}. The planets in the 1.4 – 2.6 Earth's radius range can be large rocky planets, water worlds, or having massive H/He envelopes \citep{luque2022density,rogers2023conclusive}. The figure is adapted from \cite{rogers2023conclusive} with permission.}
    \label{fig:demo}
\end{figure}

The exploration of exoplanets has revealed a staggering diversity of planets beyond the solar system. We have discovered $>6000$ validated exoplanets in our interstellar neighborhood, and most of the planets for which we have measured radii are within 3 times Earth’s radius (NASA Exoplanet Archive). In addition to a small subset of Earth-sized planets, many more discovered planets have radii in the 1.4 -- 2.6 Earth radius range, covering a planet size range not found in the solar system \citep{fulton2018california}. Current observations and models suggest that these larger-than-Earth planets can have either a predominantly rocky composition (commonly referred to as ``super-Earths'') or substantial volatile-rich layers (commonly referred to as ``sub-Neptunes''). Meanwhile, the planet demographics and the planet formation models indicate that sub-Neptunes can be either planets with massive H$_2$/He/H$_2$O envelopes or planets with a large fraction of water by mass and without any massive H/He layer \citep[sometimes referred to as ``water worlds'' or ``ocean planets''; Figure~\ref{fig:demo};][]{venturini2020nature, izidoro2022exoplanet, luque2022density, rogers2023conclusive, burn2024radius,benneke2024jwst,hu2025water}. 

It is currently unknown whether most sub-Neptunes are gas dwarfs or water worlds, and how their composition depends on bulk planetary properties and formation and evolution environments. Therefore, determining the composition of sub-Neptunes is one of the frontiers of exoplanet science today. For transiting sub-Neptunes typically in close-in orbits of stars, astronomers are pursuing their nature by obtaining precise measurements of their masses and radii, as well as measuring their atmospheric composition using HST and JWST \citep[e.g.,][]{madhusudhan2023carbon, damiano2024lhs, piaulet2024jwst}. ESA’s PLATO and ARIEL missions may detect and characterize more transiting sub-Neptunes in the next decade. 

The characterization of sub-Neptunes through transit has been and will be limited to those planets in close-in orbits. Because most rocky planets, water worlds, and gas dwarfs in the habitable zones and wider orbits of FGK stars do not transit, they will remain elusive with current facilities. The distinction between gas dwarfs and water worlds is crucial for low-temperature planets in wide orbits (i.e., the planets typically probed by the Habitable Worlds Observatory), because the temperate water worlds that have moderate-size ($<\sim50$ bars) atmospheres can host liquid-water oceans even though they are not Earth-like rocky planets \citep{goldblatt2015habitability, koll2019hot, madhusudhan2021habitability,hu2021unveiling}. It is thus essential to be able to distinguish rocky planets and water worlds from gas-rich planets for the search for habitable worlds.

This science case addresses the Questions and Discovery Areas E-Q2b, E-Q2c, and E-Q3b identified in the Astro2020 decadal survey final report.

\section{Science Objective}

Our science objective is to characterize the atmospheres and surfaces of the exoplanets discovered to the extent that distinguishes planets with massive H$_2$/He envelopes, planets with massive H$_2$O-dominated envelopes, and rocky planets with secondary atmospheres.

The current exoplanet demographics and characteristics, if extending to long-period planets (the main group of planets to be discovered by HWO), imply that HWO could detect approximately 5 times as many larger-than-Earth planets as Earth-sized planets \citep[e.g.,][]{luvoir2019luvoir}. The larger-than-Earth planets will likely include large rocky planets, water worlds, and mini Neptunes (Figure~1). Moreover, the planets to be discovered by HWO are even more likely to be water worlds than planets discovered by Kepler or TESS (and in the future by PLATO), because they are generally farther from their host stars, and thus form closer to the ice line and should accrete significant quantities of water-rich solids \citep{bitsch2021dry}. Identifying these temperate water worlds could provide a new avenue to find potentially habitable planets having liquid-water oceans. 

JWST (and ARIEL in the future) can be used to determine the nature of transiting sub-Neptune-sized exoplanets, and Extremely Large Telescopes (ELTs) on the ground may measure the reflected light spectra of non-transiting sub-Neptunes. However, most of the viable targets of these observations will be in close-in orbits of M dwarf stars with few exceptions \citep[e.g.,][]{snellen2022detecting}. To determine the atmospheric properties of sub-Neptunes in or near the habitable zones of nearby stars will require direct imaging through a mission like HWO. In the meantime, any long-period transiting sub-Neptunes found by PLATO could be preferred targets for further characterization by HWO, as the radius of these planets would be known precisely through transit measurements \citep[e.g.,][]{rauer2014plato}. If any directly imaged sub-Neptunes are accessible by both HWO and ELTs, the spectroscopic observations are likely complementary as the ELTs could cover a wavelength range redder to that of HWO.

Distinguishing rocky planets and water worlds from mini Neptunes will rely on characterizing their atmospheres. If we have perfect knowledge of a planet's radius, the rocky nature could be confirmed if the planet's mass is measured to a precision of $\sim1$ Earth mass (Figure 1). This translates to a radial-velocity precision of $\sim8$ cm s$^{-1}$ for planets in the habitable zone of the median star in the target list considered for HWO. However, observations that consistently provide such precision have not been reached yet in astronomy, and may not be possible for many stars that are mildly active or above the Kraft break \citep[e.g.,][]{crass2021extreme}. Astrometry may provide another avenue to measure the planetary mass precisely (see another SCDD on astrometry led by S.~Gaudi). More importantly, the planet's radius is degenerate with the albedo for the planets to be discovered by HWO, at least in single-band photometry. For example, a 1.7-Earth-radius planet may have the same flux as a higher-albedo 1.4-Earth-radius planet. Such a large uncertainty in the planet's radius would further complicate the determination of the rocky nature of the planet from the mass and radius alone. Meanwhile, multiband and spectroscopic measurements can improve the radius constraints through characterizing the planetary atmospheres and surfaces \citep[e.g.,][]{feng2018characterizing,damiano2021reflected}.

Mini Neptunes have H$_2$-dominated atmospheres that are at least 0.1-1\% planet mass to be consistent with the apparent planet size, corresponding to a pressure of at least $10^3 - 10^4$ bars \citep{rogers2023conclusive}. These massive atmospheres will maintain thermochemical equilibrium at depths for C, O, N, and S species \citep{fortney2020beyond,hu2021photochemistry}. By contrast, rocky planets and water worlds will have smaller atmospheres with diverse compositions, dominated by H$_2$, H$_2$O, N$_2$, CO, CO$_2$, or O$_2$, controlled by exchange with an ocean or a dry surface underneath \citep[e.g.,][]{hu2021unveiling,krissansen2021oxygen,liggins2022growth}. The composition of secondary atmospheres on rocky planets is ultimately dependent on the initial volatile inventory and planetary evolution. A rocky planet could also have volcanoes that emit sulfur gases into the atmosphere \citep{hu2013photochemistry,loftus2019sulfate} and have spectral features of lands which are distinctive from those of oceans \citep{hu2012theoretical,cowan2013determining,barrientos2023search}. Therefore, by determining the bulk atmospheric composition and measuring the abundances of key carbon-, nitrogen-, and sulfur-bearing molecules in the atmospheres of the detected exoplanets, we can characterize their intrinsic nature and identify rocky planets and water worlds. In addition to atmospheric composition, detecting surface features like ocean glint, vegetation red edge, spectral features of different land types also provides multiple lines of evidence, useful for distinguishing rocky planets and water worlds.

\section{Physical Parameters}

\begin{table*}[ht!]
    \centering
     \caption[Physical Parameters]{Physical parameters of planets to be measured to achieve this science case.}
    \label{tab:parameter}
\begin{tabularx}{480pt}{LLLLL}
        \noalign{\smallskip}
        \hline
        \noalign{\smallskip}
 \textbf{Physical Parameter} &  \textbf{State of the Art} &  \textbf{Incremental Progress} &  \textbf{Substantial Progress} &  \textbf{Major Progress}\\
        \noalign{\smallskip}
        \hline
        \noalign{\smallskip}
Presence of gases in small exoplanet atmospheres                      & JWST is detecting H$_2$O, CH$_4$, CO, CO$_2$, NH$_3$, and SO$_2$ in transiting sub-Neptunes of M stars                                                               & Detection of H$_2$O in directly imaged sub-Neptune-sized exoplanets                                        & Detection of H$_2$O, CH$_4$, and CO$_2$ in directly imaged sub-Neptune-sized exoplanets                                         & Detection of additional gases such as CO, NH$_3$, HCN, H$_2$S, and SO$_2$ in directly imaged sub-Neptune-sized exoplanets                                          \\
        \noalign{\smallskip}
        \hline
        \noalign{\smallskip}
Abundances of the gases                                               & JWST is measuring the abundances of H$_2$O, CH$_4$, CO$_2$, and SO$_2$ in transiting sub-Neptunes of M stars, sometimes to a precision of $<1.0$ dex & Measuring the mixing ratio of H$_2$O to better than 1.0 dex in sub-Neptunes in directly imaged sub-Neptunes of FGK stars & Measuring the mixing ratio of H$_2$O, CH$_4$, and CO$_2$ to better than 1.0 dex in directly imaged sub-Neptune-sized exoplanets & Measuring the mixing ratio of additional gases such as CO, NH$_3$, HCN, H$_2$S, and SO$_2$ to better than 1.0 dex in directly imaged sub-Neptune-sized exoplanets\\
        \noalign{\smallskip}
        \hline
        \noalign{\smallskip}
Dominant Atmospheric Gas                                              & JWST is determining H$_2$- versus H$_2$O-dominated atmospheres on transiting sub-Neptunes of M stars                                               & -                                                                                                           & Determining whether directly imaged sub-Neptune-sized exoplanets have H$_2$-dominated atmospheres                        & Among non-H$_2$-dominated atmospheres, determining whether they are N$_2$-, H$_2$O-, CO-, CO$_2$-, or O$_2$-dominated                                                                 \\
        \noalign{\smallskip}
        \hline
        \noalign{\smallskip}
Number of planets characterized                                       & 0                                                                                                                                            & 5                                                                                                           & 50                                                                                                                        & 100     \\ 
        \noalign{\smallskip}        
        \hline
\end{tabularx}
\end{table*}

To achieve the science objective, this science case aims to leverage HWO's capabilities to determine the dominant atmospheric gases and the abundances of key carbon-, nitrogen-, and sulfur-bearing molecules of 50 – 100 exoplanets that have a visible-wavelength planet-to-star flux ratio $<10^{-9}$. 

This science case is relevant to the planets that have a true radius ranging from 1.4 to 2.6 times Earth's radius. However, the true radius is not immediately known from direct-imaging detections and can be degenerate with the planetary albedo. By jointly fitting for the atmospheric state and radius, atmospheric retrievals can yield constraints on a planet's size. This, together with the constraints on the planet's mass (if available), will become an orthogonal measurement of the nature of the planets. Since we do not know the true radius a priori, we estimate that a 2.6-Earth-radius planet would have a planet-to-star contrast ratio at the quadrature phase of up to $\sim10^{-9}$, and therefore propose the size cutoff in terms of the planet-to-star flux ratio. This quantity is not be confused with the required instrument contrast.

We do not specify the range of relevant planet-to-star distance or equilibrium temperature for this science case. This is because, in part, the range in which a sub-Neptune could sustain liquid-water oceans is still poorly understood. For example, a sub-Neptune having an H$_2$-dominated atmosphere could host liquid-water oceans at a planet-to-star distance that is well greater than the upper limit of the habitable zone defined for rocky planets \citep{kasting1993habitable,pierrehumbert2011hydrogen}.

Due to the expected diversity of sub-Neptune-sized exoplanets, it is essential to build up a large enough sample to address population-level science questions. While initial characterization of a handful of targets could bring exciting science results on these individual planets, a sample of $\sim50$ would allow us to determine the occurrence rates of large rocky planets versus water worlds or mini-Neptunes among sub-Neptunes by $>3\sigma$ (assuming $>20\%$ in the sample turns out to be rocky planets and $>20\%$ water worlds). An enlarged sample of $\sim100$ would provide unprecedented knowledge of how the formation of large rocky planets versus water worlds or mini-Neptunes depends on macroscopic parameters such as the mass and metallicity of the host star, concurrence of giant planets, and orbital separation and architecture, potentially redefining our understanding of the formation and evolution of habitable worlds.

The expected bulk atmospheric composition ranges from H$_2$-dominated to N$_2$/H$_2$O/CO/CO$_2$/O$_2$ dominated \citep[e.g.,][]{forget2014possible,krissansen2021oxygen,lichtenberg2022geophysical}. The determination of a non-H$_2$ gas as the dominant (not just abundant) gas in the atmosphere strongly indicates the planets to be rocky planets or water worlds. If the dominant gas is found to be H$_2$, then determining whether the atmosphere has more abundant CH$_4$ or CO$_2$, as well as whether the atmosphere has abundant NH$_3$, will delineate whether the atmosphere is massive (indicating a mini Neptune) or not (indicating a rocky planet/water world) \citep{hu2021unveiling,yu2021identify,tsai2021inferring,wogan2024jwst}. If the atmosphere is found to be dominated by a non-H$_2$ gas, determining whether the atmosphere has volcanic gases such as SO$_2$ could further indicate whether the planet has a rocky surface \citep{loftus2019sulfate}. As corroborating evidence, a low-temperature water world should not have abundant NH$_3$ in the atmosphere \citep{hu2021unveiling}, although preferred partitioning of N and S in magmas could result in confounding scenarios \citep[e.g.,][]{suer2023distribution,shorttle2024}. Furthermore, understanding the internal composition of the planets based on the stellar Fe, Mg, and Si composition \citep{schulze2021probability,unterborn2023nominal,hinkel2024host} could help delineate the rocky-planet versus water-world scenarios. The literature cited in this paragraph commonly indicates that using the abundances of the trace gases to infer bulk planetary properties will require a precision of $\sim1.0$ dex in the mixing ratio, which could be used as a guiding figure subject to further refinement.

\section{Description of Observations}

\begin{figure*}[ht!]
    \centering
    \includegraphics[width=\textwidth]{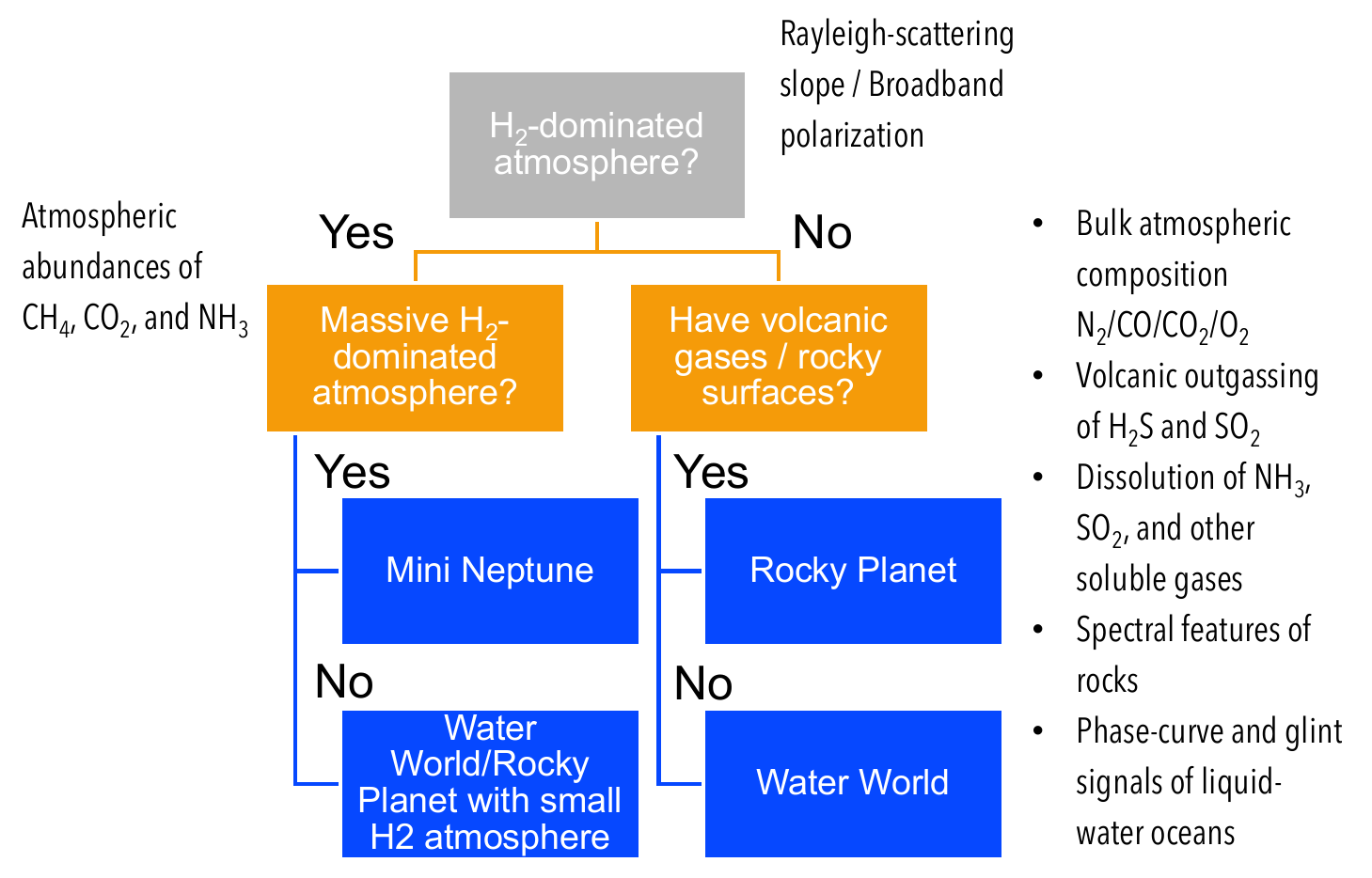}
    \caption{An observation and interpretation decision tree to characterize sub-Neptune-sized exoplanets via spectroscopy with HWO.}
    \label{fig:scheme}
\end{figure*}

\begin{table*}[ht!]
    \centering
     \caption[Observations]{Observations needed to achieve this science case.}
    \label{tab:observations}
\begin{tabularx}{480pt}{LLLLL}
        \noalign{\smallskip}
        \hline
        \noalign{\smallskip}
 \textbf{Observations Needed} &  \textbf{State of the Art} &  \textbf{Incremental Progress} &  \textbf{Substantial Progress} &  \textbf{Major Progress}\\
        \noalign{\smallskip}
        \hline
        \noalign{\smallskip}
Reflectance spectroscopy of exoplanets with a planet-to-star flux ratio \textless{}$10^{-9}$                     & - & Spectroscopy in $0.4 - 1.0\ \mu$m with R\textgreater{}140 and SNR\textgreater{}10 & Spectroscopy in $0.4 - 1.0\ \mu$m with R\textgreater{}140 and SNR\textgreater{}20, and spectroscopy in $1.0 - 1.7\ \mu$m with R\textgreater{}40 and SNR\textgreater{}10 & Spectroscopy in $0.4 - 1.0\ \mu$m with R\textgreater{}140 and SNR\textgreater{}40, and spectroscopy in $1.0 - 1.7\ \mu$m with R\textgreater{}70 (to be refined) and SNR\textgreater{}10, and spectroscopy in $0.25 - 0.4\ \mu$m with R\textgreater{}7 and SNR\textgreater{}10 \\
        \noalign{\smallskip}
        \hline
        \noalign{\smallskip}
Broadband polarimetry in the reflected light of exoplanets with a planet-to-star flux ratio \textless{}$10^{-9}$ & - & Single-band polarimetry with a precision of $\pm5\%$ at 2-3 epochs            & Single-band polarimetry with a precision of $\pm3\%$ at 3-5 epochs                                                                                             & Multi-band polarimetry with a precision of $\pm1\%$ at 5-10 epochs                                                                                                                                                                                                              \\
        \noalign{\smallskip}
        \hline
        \noalign{\smallskip}
Number of planets characterized with a planet-to-star flux ratio \textless{}$10^{-9}$                                                                           & - & 5                                                                            & 50                                                                                                                                                            & 100                                                                                                                                                                                                                                                                            \\
        \noalign{\smallskip}
        \hline
        \noalign{\smallskip}
Measuring the mass of the planets (mass ranging in 3 - 20 Earth mass)                                       & - & -                                                                            & Better than $3\sigma$                                                                                                                                                & Better than $10\sigma$      \\                                                                                                                                                                                                                                                         
                                                             \noalign{\smallskip}
        \hline
\end{tabularx}
\end{table*}

To achieve the measurements outlined above, HWO will need to conduct observations designed to measure the reflectance spectra of 50 – 100 exoplanets with a visible-wavelength planet-to-star flux ratio $<10^{-9}$ from 0.25 to 1.7 $\mu$m, at a spectral resolution of $R>140$, and SNR$>$10 per spectral element. These observations will leverage HWO's advanced capabilities in high-contrast imaging.

To detect a single gas (such as H$_2$O), spectroscopy in narrow ($\sim20\%$) bands strategically placed in $0.6 - 1.0\ \mu$m, with a spectral resolution of $>70$ and SNR$>$10 per spectral element (defined at the continuum), would suffice \citep{latouf2023bayesian,young2024retrievals}. However, narrowband spectroscopy cannot precisely measure the mixing ratio of the gas due to the lack of sufficient out-of-band baseline and overall understanding of the context. Measuring the mixing ratio of H$_2$O to better than 1.0 dex and to distinguish a H$_2$O-rich atmosphere from a H$_2$O-dominated one will require spectroscopy that covers a broader wavelength range ($0.4 - 1.0\ \mu$m) and with a higher spectral resolution of $R>140$ \citep{feng2018characterizing,damiano2021reflected,damiano2022reflected}.

Because H$_2$ itself does not absorb significantly in the wavelength bands expected to be covered by HWO \citep[except, perhaps, the weak collision-induced opacity near $\sim1.2\ \mu$m,][]{koroleva2024collision}, to further determine whether the atmosphere is H$_2$-dominated or not could rely on (1) tallying the abundances of the main radiatively active components of the atmosphere (e.g., H$_2$O, CO$_2$), (2) inference from the chemical signature of the atmosphere (e.g., the CH$_4$-to-CO$_2$ ratio), and (3) inference from the Rayleigh scattering slope of the reflectance spectra. Method (1) is proven to be very hard because the reflectance spectra often lose dependence on the exact mixing ratio when the mixing ratio is above $\sim1\%$ for H$_2$O \citep{damiano2021reflected}. We thus propose to aim at both (2) and (3) as they are complementary to each other. Method (2) would require measuring the mixing ratios of CH$_4$ and CO$_2$ to better than 1.0 dex, and this can be achieved by expanding the spectroscopy to the near-infrared wavelength range ($1.0 - 1.7\ \mu$m) with a spectral resolution $R>40$ and SNR$>$10 per spectral element \citep{damiano2022reflected}. Method (3) could be achieved by measuring the reflectance spectrum precisely in visible light \citep[SNR$>$20,][]{hall2023constraining}.

If a planet is found to not have a massive H$_2$ atmosphere, this opens up opportunities to further characterize its nature. (1) Is its atmosphere dominated by N$_2$, CO, CO$_2$, or O$_2$? (2) Does it have volcanic outgassing of H$_2$S and SO$_2$ as signposts of rocky planets? (3) Does it have a liquid-water ocean that can regulate the atmospheric abundance of CO$_2$ and dissolve any soluble gases such as NH$_3$, HCN, and SO$_2$? For (1), it has been shown that an SNR$\sim$40 in visible light is necessary to reliably break the degeneracy between an O$_2$-dominated versus an CO/N$_2$-dominated atmosphere, and because CO and N$_2$ have similar molecular weights, it is also necessary to observe the CO absorption feature at 1.6 $\mu$m \citep{hall2023constraining}. The O$_2$-O$_2$ collision-induced opacities in the near-infrared could help break this degeneracy for atmospheres larger than a few bars \citep[e.g.,][]{leung2020high}. For (2) and (3), low-resolution spectroscopy in the UV band ($0.25 - 0.4\ \mu$m) can provide sensitivity to H$_2$S and SO$_2$ and their photochemical products, in addition to O$_3$ \citep{hu2013photochemistry,gao2017sulfur,damiano2023reflected}. The detectability of NH$_3$ has not been assessed specifically for reflectance spectroscopy, but it has substantial absorption bands in $1.5 - 1.7\ \mu$m. However, NH$_3$, CH$_4$, H$_2$O, CO, and CO$_2$ all have spectral features in the narrow wavelength range between 1.4 and 1.8 $\mu$m, and thus a higher spectral resolution in the near-infrared band is likely necessary to measure their mixing ratios. In addition, liquid and ice water clouds imprint characteristic signatures on the reflected light spectra in the near-infrared, particularly by modulating the continuum, which can be used to constrain cloud properties \citep[e.g.,][]{damiano2022reflected,kofman2024pale,roccetti2025planet}. Direct detection of liquid-water oceans can be made by measuring the planetary phase curves and ocean glint \citep[e.g.][]{cowan2009alien,robinson2010detecting,lustig2018detecting,vaughan2023chasing}. Relevant measurements for direct detection of liquid-water oceans are discussed in a separate SCDD developed by N.~Cowan, J.~Lustig-Yaeger, and collaborators.

In addition, to break the degeneracy between an O$_2$-dominated versus an N$_2$- or CO-dominated atmosphere using the Rayleigh-scattering slope in the reflected light spectrum, it is necessary to have strong constraints on the planetary mass independently from the reflectance spectrum \citep{damiano2025effects} and achieve a high SNR (40) per spectral element in the visible wavelengths \citep{hall2023constraining}. It has been shown recently that a mass uncertainty of 10\% would enable the spectral characterization of the bulk atmospheric composition \citep{damiano2025effects}. See the SCDD developed by S.~Gaudi and collaborators for an analysis of the feasibility of achieving such high-precision mass constraints via astrometry.

If a temperate planet is found to be a rocky planet or water world, it can conceivably host a liquid-water ocean. Then, the observations and measurements laid out in the previously-mentioned ocean detection SCDD should be applied, with the modification of the planet-to-star contrast ratio that could be as large as $10^{-9}$ in the visible band. We note that the actual flow of observations might differ from what is presented in this document, e.g., detecting a liquid-water ocean through glint or phase-curve signals might be easier than detailed spectroscopy for a subset of candidate planets. We identify this point to be an essential area for further study and optimization.

In addition, distinguishing between a planet with lands versus a planet fully covered by the ocean could also leverage the typical spectral features of lands in the visible and NIR bands \citep{hu2012theoretical,barrientos2023search}. For candidate rocky planets, detecting spectral features of typical rocks (e.g., basalts, quartz, iron oxides, etc) provides a direct confirmation of their rocky nature. For reflectance spectroscopy with HWO, the most relevant spectral features come from electronic transitions of F$^{2+}$ in the crystal fields of mafic rocks (such as olivine and pyroxene, absorbing broadly near 1 $\mu$m and 2 $\mu$m), typical absorption bands of bounded hydroxyl and water of hydrated rocks (at 1.4 $\mu$m and 1.9 $\mu$m), as well as reflectance slopes due to space weathering and oxidative weathering \citep{hu2012theoretical}. The rock features and atmospheric features are potentially distinguishable thanks to their different spectral shape. Further studies are necessary to determine the exact spectroscopic measurements needed to achieve this feat.

Last but not least, polarimetry will be an invaluable tool in helping to distinguish between different types of planetary atmospheres and surfaces. In the 600--800 nm range, the peak degree of linear polarization can range from small values ($<10\%$) for a rocky surface or a cloudy atmosphere, to very large values ($>40\%$) for a Rayleigh scattering atmosphere. This wide range of conditions have readily been seen in the solar system in disk-integrated measurements and models: 6-10\% for Moon, Mars, and Mercury, 30-50\% for Neptune and Titan (in the Rayleigh-dominated portion of their spectra), and typically 5--10\% for other giant planets and Venus. Earth, with its complex mix of a liquid surface, rocky ground, ice, vegetation, and a Rayleigh scattering atmosphere uniquely has a polarization in the 15-20\% range \citep[e.g.,][]{gordon2023polarized}. Notably, orbital phase-resolved polarimetric measurements of Venus were used to identify sulfuric acid as the prime constituent of Venus's clouds \citep{hansen1974interpretation}. 

While the polarization signal of the glint feature of a planet covered by liquid-water oceans is addressed in the separate ocean detection SCDD, broadband polarimetric measurements will more generally help distinguish the various types of small exoplanets laid out in this SCDD. Polarimetry is complementary to spectroscopy to diagnose the surface/atmospheric parameters because the polarimetric signals discussed above are all broadband features mostly in the visible wavelengths. In order to distinguish between the Rayleigh scattering dominated scenarios and an Earth-twin, a precision on the degree of polarization of $\sim\pm5\%$ would facilitate a $3\sigma$ distinction of the peak polarizations. In order to distinguish between an Earth-twin and the Mie-scattering and regolith scenarios, a precision of $\sim\pm3\%$ would be needed. Broadband polarimetric measurements will also need to be obtained at several phase angles to infer (or directly measure) the peak polarization.

We emphasize that the observation characteristics laid out in this section and in Table~\ref{tab:observations} are based on solar system experiences and recent studies on reflectance spectroscopy of Earth-sized rocky planets \citep[with the exception of][]{damiano2021reflected}. The figures for wavelength coverage, spectral resolution, and SNR are subject to future refinement. It is necessary to carry out studies of the reflectance spectroscopy that focuses on sub-Neptune-sized exoplanets, including and differentiating large rocky planets, water worlds, and mini-Neptunes, to optimize and pinpoint the observations needed to achieve this science case.

{\bf Acknowledgements.} Part of the research was carried out at the Jet Propulsion Laboratory, California Institute of Technology, under a contract with the National Aeronautics and Space Administration.


\begin{thebibliography}{}
        \parskip=0pt \itemsep=0pt \small \baselineskip=11pt
        \expandafter\ifx\csname natexlab\endcsname\relax\def\natexlab#1{#1}\fi
        \providecommand{\url}[1]{\href{#1}{#1}}
        \providecommand{\dodoi}[1]{}
        \providecommand{\doeprint}[1]{\href{http://ascl.net/#1}{#1}}
        \providecommand{\doarXiv}[1]{\href{https://arxiv.org/abs/#1}{arXiv:#1}}
        
        \bibitem[{Barrientos {et~al.}(2023)Barrientos, MacDonald, Lewis, \& Kaltenegger}]{barrientos2023search}
        Barrientos, J.~G., MacDonald, R.~J., Lewis, N.~K., {et~al.} 2023, The Astrophysical Journal, 946, 96
        
        \bibitem[{Benneke {et~al.}(2024)Benneke, Roy, Coulombe, Radica, Piaulet, Ahrer, Pierrehumbert, Krissansen-Totton, Schlichting, Hu, {et~al.}}]{benneke2024jwst}
        Benneke, B., Roy, P.-A., Coulombe, L.-P., {et~al.} 2024, arXiv preprint arXiv:2403.03325
        
        \bibitem[{Bitsch {et~al.}(2021)Bitsch, Raymond, Buchhave, Bello-Arufe, Rathcke, \& Schneider}]{bitsch2021dry}
        Bitsch, B., Raymond, S.~N., Buchhave, L.~A., {et~al.} 2021, Astronomy \& Astrophysics, 649, L5
        
        \bibitem[{Burn {et~al.}(2024)Burn, Mordasini, Mishra, Haldemann, Venturini, Emsenhuber, \& Henning}]{burn2024radius}
        Burn, R., Mordasini, C., Mishra, L., {et~al.} 2024, Nature astronomy, 8, 463
        
        \bibitem[{Cowan \& Strait(2013)}]{cowan2013determining}
        Cowan, N.~B., \& Strait, T.~E. 2013, The Astrophysical Journal Letters, 765, L17
        
        \bibitem[{Cowan {et~al.}(2009)Cowan, Agol, Meadows, Robinson, Livengood, Deming, Lisse, A'Hearn, Wellnitz, Seager, {et~al.}}]{cowan2009alien}
        Cowan, N.~B., Agol, E., Meadows, V.~S., {et~al.} 2009, The Astrophysical Journal, 700, 915
        
        \bibitem[{Crass {et~al.}(2021)Crass, Gaudi, Leifer, Beichman, Bender, Blackwood, Burt, Callas, Cegla, Diddams, {et~al.}}]{crass2021extreme}
        Crass, J., Gaudi, B.~S., Leifer, S., {et~al.} 2021, arXiv preprint arXiv:2107.14291
        
        \bibitem[{Damiano {et~al.}(2024)Damiano, Bello-Arufe, Yang, \& Hu}]{damiano2024lhs}
        Damiano, M., Bello-Arufe, A., Yang, J., {et~al.} 2024, The Astrophysical Journal Letters, 968, L22
        
        \bibitem[{Damiano {et~al.}(2025)Damiano, Burr, Hu, Burt, \& Kataria}]{damiano2025effects}
        Damiano, M., Burr, Z., Hu, R., {et~al.} 2025, The Astronomical Journal, 169, 97
        
        \bibitem[{Damiano \& Hu(2021)}]{damiano2021reflected}
        Damiano, M., \& Hu, R. 2021, The Astronomical Journal, 162, 200
        
        \bibitem[{Damiano \& Hu(2022)}]{damiano2022reflected}
        ---. 2022, The Astronomical Journal, 163, 299
        
        \bibitem[{Damiano {et~al.}(2023)Damiano, Hu, \& Mennesson}]{damiano2023reflected}
        Damiano, M., Hu, R., \& Mennesson, B. 2023, The Astronomical Journal, 166, 157
        
        \bibitem[{Feng {et~al.}(2018)Feng, Robinson, Fortney, Lupu, Marley, Lewis, Macintosh, \& Line}]{feng2018characterizing}
        Feng, Y.~K., Robinson, T.~D., Fortney, J.~J., {et~al.} 2018, The Astronomical Journal, 155, 200
        
        \bibitem[{Forget \& Leconte(2014)}]{forget2014possible}
        Forget, F., \& Leconte, J. 2014, Philosophical Transactions of the Royal Society A: Mathematical, Physical and Engineering Sciences, 372, 20130084
        
        \bibitem[{Fortney {et~al.}(2020)Fortney, Visscher, Marley, Hood, Line, Thorngren, Freedman, \& Lupu}]{fortney2020beyond}
        Fortney, J.~J., Visscher, C., Marley, M.~S., {et~al.} 2020, The Astronomical Journal, 160, 288
        
        \bibitem[{Fulton \& Petigura(2018)}]{fulton2018california}
        Fulton, B.~J., \& Petigura, E.~A. 2018, The Astronomical Journal, 156, 264
        
        \bibitem[{Gao {et~al.}(2017)Gao, Marley, Zahnle, Robinson, \& Lewis}]{gao2017sulfur}
        Gao, P., Marley, M.~S., Zahnle, K., {et~al.} 2017, The Astronomical Journal, 153, 139
        
        \bibitem[{Goldblatt(2015)}]{goldblatt2015habitability}
        Goldblatt, C. 2015, Astrobiology, 15, 362
        
        \bibitem[{Gordon {et~al.}(2023)Gordon, Karalidi, Bott, Miles-P{\'a}ez, Mulder, \& Stam}]{gordon2023polarized}
        Gordon, K.~E., Karalidi, T., Bott, K.~M., {et~al.} 2023, The Astrophysical Journal, 945, 166
        
        \bibitem[{Hall {et~al.}(2023)Hall, Krissansen-Totton, Robinson, Salvador, \& Fortney}]{hall2023constraining}
        Hall, S., Krissansen-Totton, J., Robinson, T., {et~al.} 2023, The Astronomical Journal, 166, 254
        
        \bibitem[{Hansen \& Hovenier(1974)}]{hansen1974interpretation}
        Hansen, J.~E., \& Hovenier, J. 1974, J. atmos. Sci, 31, 1137
        
        \bibitem[{Hinkel {et~al.}(2024)Hinkel, Youngblood, \& Soares-Furtado}]{hinkel2024host}
        Hinkel, N.~R., Youngblood, A., \& Soares-Furtado, M. 2024, Reviews in Mineralogy and Geochemistry, 90, 1
        
        \bibitem[{Hu(2021)}]{hu2021photochemistry}
        Hu, R. 2021, The Astrophysical Journal, 921, 27
        
        \bibitem[{Hu {et~al.}(2021)Hu, Damiano, Scheucher, Kite, Seager, \& Rauer}]{hu2021unveiling}
        Hu, R., Damiano, M., Scheucher, M., {et~al.} 2021, The Astrophysical Journal Letters, 921, L8
        
        \bibitem[{Hu {et~al.}(2012)Hu, Ehlmann, \& Seager}]{hu2012theoretical}
        Hu, R., Ehlmann, B.~L., \& Seager, S. 2012, The Astrophysical Journal, 752, 7
        
        \bibitem[{Hu {et~al.}(2013)Hu, Seager, \& Bains}]{hu2013photochemistry}
        Hu, R., Seager, S., \& Bains, W. 2013, The Astrophysical Journal, 769, 6
        
        \bibitem[{Hu {et~al.}(2025)Hu, Bello-Arufe, Tokadjian, Yang, Damiano, Roy, Coulombe, Madhusudhan, Constantinou, \& Benneke}]{hu2025water}
        Hu, R., Bello-Arufe, A., Tokadjian, A., {et~al.} 2025, arXiv preprint arXiv:2507.12622
        
        \bibitem[{Izidoro {et~al.}(2022)Izidoro, Schlichting, Isella, Dasgupta, Zimmermann, \& Bitsch}]{izidoro2022exoplanet}
        Izidoro, A., Schlichting, H.~E., Isella, A., {et~al.} 2022, The Astrophysical Journal Letters, 939, L19
        
        \bibitem[{Kasting {et~al.}(1993)Kasting, Whitmire, \& Reynolds}]{kasting1993habitable}
        Kasting, J.~F., Whitmire, D.~P., \& Reynolds, R.~T. 1993, Icarus, 101, 108
        
        \bibitem[{Kofman {et~al.}(2024)Kofman, Villanueva, Fauchez, Mandell, Johnson, Payne, Latouf, \& Kelkar}]{kofman2024pale}
        Kofman, V., Villanueva, G.~L., Fauchez, T.~J., {et~al.} 2024, The Planetary Science Journal, 5, 197
        
        \bibitem[{Koll \& Cronin(2019)}]{koll2019hot}
        Koll, D.~D., \& Cronin, T.~W. 2019, The Astrophysical Journal, 881, 120
        
        \bibitem[{Koroleva {et~al.}(2024)Koroleva, Kassi, Fleurbaey, \& Campargue}]{koroleva2024collision}
        Koroleva, A., Kassi, S., Fleurbaey, H., {et~al.} 2024, Journal of Quantitative Spectroscopy and Radiative Transfer, 318, 108948
        
        \bibitem[{Krissansen-Totton {et~al.}(2021)Krissansen-Totton, Fortney, Nimmo, \& Wogan}]{krissansen2021oxygen}
        Krissansen-Totton, J., Fortney, J.~J., Nimmo, F., {et~al.} 2021, AGU Advances, 2, e2020AV000294
        
        \bibitem[{Latouf {et~al.}(2023)Latouf, Mandell, Villanueva, Moore, Susemiehl, Kofman, \& Himes}]{latouf2023bayesian}
        Latouf, N., Mandell, A.~M., Villanueva, G.~L., {et~al.} 2023, The Astronomical Journal, 166, 129
        
        \bibitem[{Leung {et~al.}(2020)Leung, Meadows, \& Lustig-Yaeger}]{leung2020high}
        Leung, M., Meadows, V.~S., \& Lustig-Yaeger, J. 2020, The Astronomical Journal, 160, 11
        
        \bibitem[{{Lichtenberg} {et~al.}(2023){Lichtenberg}, {Schaefer}, {Nakajima}, \& {Fischer}}]{lichtenberg2022geophysical}
        {Lichtenberg}, T., {Schaefer}, L.~K., {Nakajima}, M., {et~al.} 2023, Protostars and Planets VII, 534, 907
        
        \bibitem[{Liggins {et~al.}(2022)Liggins, Jordan, Rimmer, \& Shorttle}]{liggins2022growth}
        Liggins, P., Jordan, S., Rimmer, P.~B., {et~al.} 2022, Journal of Geophysical Research: Planets, 127, e2021JE007123
        
        \bibitem[{Loftus {et~al.}(2019)Loftus, Wordsworth, \& Morley}]{loftus2019sulfate}
        Loftus, K., Wordsworth, R.~D., \& Morley, C.~V. 2019, The Astrophysical Journal, 887, 231
        
        \bibitem[{Luque \& Pall{\'e}(2022)}]{luque2022density}
        Luque, R., \& Pall{\'e}, E. 2022, Science, 377, 1211
        
        \bibitem[{Lustig-Yaeger {et~al.}(2018)Lustig-Yaeger, Meadows, Mendoza, Schwieterman, Fujii, Luger, \& Robinson}]{lustig2018detecting}
        Lustig-Yaeger, J., Meadows, V.~S., Mendoza, G.~T., {et~al.} 2018, The Astronomical Journal, 156, 301
        
        \bibitem[{{LUVOIR Team} {et~al.}(2019)}]{luvoir2019luvoir}
        {LUVOIR Team}, {et~al.} 2019, arXiv preprint arXiv:1912.06219
        
        \bibitem[{Madhusudhan {et~al.}(2021)Madhusudhan, Piette, \& Constantinou}]{madhusudhan2021habitability}
        Madhusudhan, N., Piette, A.~A., \& Constantinou, S. 2021, The Astrophysical Journal, 918, 1
        
        \bibitem[{Madhusudhan {et~al.}(2023)Madhusudhan, Sarkar, Constantinou, Holmberg, Piette, \& Moses}]{madhusudhan2023carbon}
        Madhusudhan, N., Sarkar, S., Constantinou, S., {et~al.} 2023, The Astrophysical Journal Letters, 956, L13
        
        \bibitem[{Parc {et~al.}(2024)Parc, Bouchy, Venturini, Dorn, \& Helled}]{parc2024super}
        Parc, L., Bouchy, F., Venturini, J., {et~al.} 2024, Astronomy \& Astrophysics, 688, A59
        
        \bibitem[{Piaulet-Ghorayeb {et~al.}(2024)Piaulet-Ghorayeb, Benneke, Radica, Raul, Coulombe, Ahrer, Kubyshkina, Howard, Krissansen-Totton, MacDonald, {et~al.}}]{piaulet2024jwst}
        Piaulet-Ghorayeb, C., Benneke, B., Radica, M., {et~al.} 2024, The Astrophysical Journal Letters, 974, L10
        
        \bibitem[{Pierrehumbert \& Gaidos(2011)}]{pierrehumbert2011hydrogen}
        Pierrehumbert, R., \& Gaidos, E. 2011, The Astrophysical Journal Letters, 734, L13
        
        \bibitem[{Rauer {et~al.}(2014)Rauer, Catala, Aerts, Appourchaux, Benz, Brandeker, Christensen-Dalsgaard, Deleuil, Gizon, Goupil, {et~al.}}]{rauer2014plato}
        Rauer, H., Catala, C., Aerts, C., {et~al.} 2014, Experimental Astronomy, 38, 249
        
        \bibitem[{Robinson {et~al.}(2010)Robinson, Meadows, \& Crisp}]{robinson2010detecting}
        Robinson, T.~D., Meadows, V.~S., \& Crisp, D. 2010, The Astrophysical Journal Letters, 721, L67
        
        \bibitem[{Roccetti {et~al.}(2025)Roccetti, Emde, Sterzik, Manev, Seidel, \& Bagnulo}]{roccetti2025planet}
        Roccetti, G., Emde, C., Sterzik, M.~F., {et~al.} 2025, Astronomy \& Astrophysics, 697, A170
        
        \bibitem[{Rogers {et~al.}(2023)Rogers, Schlichting, \& Owen}]{rogers2023conclusive}
        Rogers, J.~G., Schlichting, H.~E., \& Owen, J.~E. 2023, The Astrophysical Journal Letters, 947, L19
        
        \bibitem[{Schulze {et~al.}(2021)Schulze, Wang, Johnson, Gaudi, Unterborn, \& Panero}]{schulze2021probability}
        Schulze, J., Wang, J., Johnson, J., {et~al.} 2021, The Planetary Science Journal, 2, 113
        
        \bibitem[{{Shorttle} {et~al.}(2024){Shorttle}, {Jordan}, {Nicholls}, {Lichtenberg}, \& {Bower}}]{shorttle2024}
        {Shorttle}, O., {Jordan}, S., {Nicholls}, H., {et~al.} 2024, The Astrophysical Journal Letters, 962, L8
        
        \bibitem[{Snellen {et~al.}(2022)Snellen, Snik, Kenworthy, Albrecht, Anglada-Escud{\'e}, Baraffe, Baudoz, Benz, Beuzit, Biller, {et~al.}}]{snellen2022detecting}
        Snellen, I.~A., Snik, F., Kenworthy, M., {et~al.} 2022, Experimental Astronomy, 54, 1237
        
        \bibitem[{Suer {et~al.}(2023)Suer, Jackson, Grewal, Dalou, \& Lichtenberg}]{suer2023distribution}
        Suer, T.-A., Jackson, C., Grewal, D.~S., {et~al.} 2023, Frontiers in Earth Science, 11, 1159412
        
        \bibitem[{Tsai {et~al.}(2021)Tsai, Innes, Lichtenberg, Taylor, Malik, Chubb, \& Pierrehumbert}]{tsai2021inferring}
        Tsai, S.-M., Innes, H., Lichtenberg, T., {et~al.} 2021, The Astrophysical Journal Letters, 922, L27
        
        \bibitem[{Unterborn {et~al.}(2023)Unterborn, Desch, Haldemann, Lorenzo, Schulze, Hinkel, \& Panero}]{unterborn2023nominal}
        Unterborn, C.~T., Desch, S.~J., Haldemann, J., {et~al.} 2023, The Astrophysical Journal, 944, 42
        
        \bibitem[{Vaughan {et~al.}(2023)Vaughan, Gebhard, Bott, Casewell, Cowan, Doelman, Kenworthy, Mazoyer, Millar-Blanchaer, Trees, {et~al.}}]{vaughan2023chasing}
        Vaughan, S.~R., Gebhard, T.~D., Bott, K., {et~al.} 2023, Monthly Notices of the Royal Astronomical Society, 524, 5477
        
        \bibitem[{Venturini {et~al.}(2020)Venturini, Guilera, Haldemann, Ronco, \& Mordasini}]{venturini2020nature}
        Venturini, J., Guilera, O.~M., Haldemann, J., {et~al.} 2020, Astronomy \& Astrophysics, 643, L1
        
        \bibitem[{Wogan {et~al.}(2024)Wogan, Batalha, Zahnle, Krissansen-Totton, Tsai, \& Hu}]{wogan2024jwst}
        Wogan, N.~F., Batalha, N.~E., Zahnle, K.~J., {et~al.} 2024, The Astrophysical Journal Letters, 963, L7
        
        \bibitem[{Young {et~al.}(2024)Young, Crouse, Arney, Domagal-Goldman, Robinson, \& Bastelberger}]{young2024retrievals}
        Young, A.~V., Crouse, J., Arney, G., {et~al.} 2024, The Planetary Science Journal, 5, 7
        
        \bibitem[{Yu {et~al.}(2021)Yu, Moses, Fortney, \& Zhang}]{yu2021identify}
        Yu, X., Moses, J.~I., Fortney, J.~J., {et~al.} 2021, The Astrophysical Journal, 914, 38
        
        \end{thebibliography}
\end{document}